\documentclass[conference,letterpaper]{IEEEtran}
\usepackage{amsfonts}
\usepackage{geometry}
\IEEEoverridecommandlockouts

\ifCLASSINFOpdf
\else
\fi
\usepackage{epstopdf}
\usepackage{multirow}
\usepackage{cite}
\usepackage{stfloats}
\usepackage{color}
\usepackage{epsfig}
\usepackage{graphicx}
\usepackage{psfig}
\usepackage{epsf}
\usepackage{slashbox}
\usepackage{float}
\usepackage{amssymb }
\usepackage[cmex10]{amsmath}
\usepackage{algorithm}
\usepackage{algorithmic}
\usepackage{booktabs}
\usepackage{amsmath,bm}
\usepackage{amsmath,amsfonts}
\usepackage[justification=centering]{caption}
\begin{document}
\title{Is the Performance of NOMA-aided Integrated Sensing and Multicast-Unicast Communications Improved by IRS?}
\author{Yang Gou,   Yinghui Ye, Guangyue Lu,  Lu Lv, and Rose Qingyang Hu
\thanks{This work was supported by  the National Natural Science Foundation of China under Grant 62201451 and the Young Talent fund of University Association for Science and Technology in Shaanxi under Grant 20210121. (\emph{Corresponding Author: Yinghui Ye})}
\thanks{Yang Gou, Yinghui Ye and Guangyue Lu are with the Shaanxi Key Laboratory of Information Communication Network and Security, Xi'an University of Posts \& Telecommunications, China (e-mail: evegy@stu.xupt.edu.cn, connectyyh@126.com, tonylugy@163.com).}
\thanks{Lu Lv is with the School of Telecommunications Engineering, Xidian University, Xi'an 710071, China (e-mail: lulv@xidian.edu.cn).}
\thanks{Rose Qingyang Hu is with the Department of Electrical and Computer
Engineering at Utah State University, U.S.A. (e-mail: rose.hu@usu.edu).}
}
\markboth{}
{Shi\MakeLowercase{\textit{et al.}}:}
\maketitle
\begin{abstract}
In this paper, we consider intelligent reflecting surface (IRS) in a non-orthogonal multiple access (NOMA)-aided Integrated Sensing and Multicast-Unicast Communication (ISMUC) system, where the multicast signal is used for sensing and communications while the unicast signal is used only for communications. Our goal is to depict whether the IRS improves the performance of NOMA-ISMUC system or not under the imperfect/perfect successive interference cancellation (SIC) scenario. Towards this end, we formulate a non-convex problem to maximize the unicast rate while ensuring the minimum target illumination power and multicast rate. To settle this problem, we employ the Dinkelbach method to transform this original problem into an equivalent one, which is then solved via  alternating optimization algorithm and semidefinite relaxation (SDR) with Sequential Rank-One Constraint Relaxation (SROCR).
Based on this, an iterative algorithm is devised to obtain a near-optimal solution.
Computer simulations verify the quick convergence of the devised iterative algorithm, and provide insightful results. Compared to NOMA-ISMUC without IRS, IRS-aided NOMA-ISMUC achieves a higher rate with perfect SIC but keeps the almost same rate in the case of imperfect SIC.
\end{abstract}
\IEEEpeerreviewmaketitle
\section{Introduction}
As a crucial technology of the sixth-generation (6G), integrated sensing and communication (ISAC) is able to perform wireless sensing and communication in the same resource and facility\cite{9839026}. In previous ISAC with multiple communication users (C-users), it was assumed that C-users operate in the orthogonal multiple access (OMA) manner, leaving a room for spectrum efficiency improvement. It has been widely assumed that non-orthogonal multiple access (NOMA) achieves a larger spectrum efficiency than OMA\cite{9345507}, which therefore motivates us using non-orthogonal multiple access (NOMA) to replace OMA in ISAC, yielding a concept called non-orthogonal ISAC (NISAC).

In recent years, there have been several contributions regarding NISAC. The authors in \cite{9729087} considered a NISAC network, where the base station (BS) transmits the mixed multicast and unicast messages, and maximizes the unicast rate subject to the minimum multicast rate constraints. The authors in \cite{9668964} proposed a beamforming design to maximize the weighted sum of the communication throughput and the effective sensing power while ensuring the similar levels of sensing power in different target directions. Considering the influence of both inter-user interference and Rayleigh channel fading, the authors in \cite{5454545} derived the closed-form expressions of the outage probability, the ergodic communication rate, and the sensing rate. Taking secure communications into account, the authors in \cite{9854898} proposed a NOMA-based secure transmission beamforming method to maximize the radar-only beamforming matrix under a given transmit power.

Despite the recent contributions, NISAC still faces challenges particularly caused by the wireless fading. Owing to the ability to intelligently control the wireless channel by adjusting the phase shift of reflection components\cite{9416177}, intelligent reflecting surface (IRS) has been recently studied in NISAC in\cite{1111333,1111555,9240028,1111444}.
The authors in \cite{1111333} proposed an IRS-NISAC system, where the dedicated IRS creates virtual LoS links for the radar targets, and maximizes beam-pattern gain by ensuring the minimum quality of service (QoS) requirement.
The paper in\cite{1111555} considered a distributed IRS assisted concurrent communication and location sensing for a blind-zone user, and minimized the Cramer-Rao lower bound to characterize the performances of both communication and location sensing in a unified manner by optimizing the active and passive beamforming.
The authors in\cite{9240028} proposed to minimize the transmission power via jointly optimizing the beamforming vectors and the IRS phase shift matrix under the QoS constraints.
In\cite{1111444}, the authors maximized the minimum radar beam-pattern gain by jointly optimizing the active beamforming, power allocation, and passive beamforming under the constraints of the minimum QoS requirement.

We note that the communication performance has not been well investigated in an IRS-NISAC network.
In practice, the communication rate is also an important performance metric in IRS-NISAC, thus, it is of significance to investigate the communication rate maximization problem subject to the minimum radar beam-pattren gain. Besides, all the existing works \cite{9729087,9668964,5454545,9854898,9416177,1111333,1111555,9240028,1111444} assumed a perfect SIC for NOMA transmissions, which does not hold in practical communications. Recall that in IRS-NISAC with imperfect SIC, IRS amplifies the NOMA signals simultaneously, that is to say, it also boosts the co-channel interference cased by the NOMA. In this case, the following question arises: is the communication performance of NISAC really improved by IRS under the imperfect SIC scenario? However, such a question has not been explicitly answered in the existing literature.

In this paper, we consider an IRS aided NOMA-Integrated Sensing and Multicast-Unicast Communication
(ISMUC) network consisting of one BS, one near-user (NU) and one far-user (FU). In particular, the BS transmits the multicast-unicast signal to NU and FU for sensing and communications. We are interested in maximizing  the unicast communication rate at the NU under the target illumination power constraint in both perfect/imperfect SIC scenarios. Our main contributions are listed below.
\begin{figure} [t]
  \centering
  \includegraphics[width=0.37\textwidth]{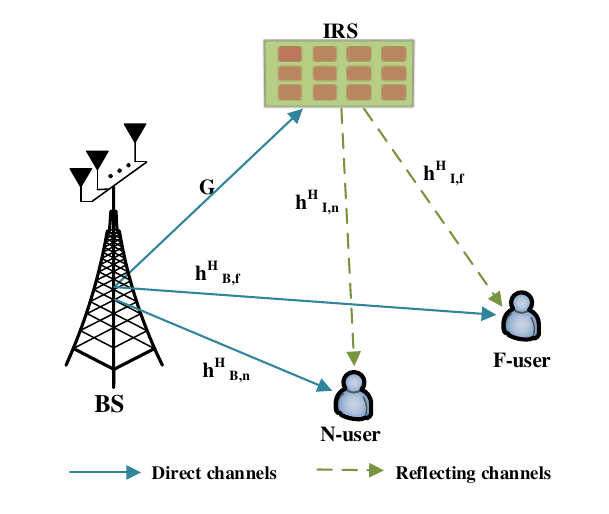}\\
  \caption{System model.}\label{fig1}
\end{figure}
\begin{itemize}
\item Considering the imperfect SIC at NU, we formulate a nonlinear programming problem to maximize the unicast rate under the constraint of the multicast rate requirement and the minimum target illumination power.  \emph{Note that the perfect SIC is a special case of the imperfect SIC, therefore, our formulated problem can also be applicable for the  perfect SIC case.}
\item To solve the nonlinear programming problem, we first employ the Dinklbach method to create an equivalent objective function of the original problem. Subsequently, we adopt the alternating optimization algorithm to decompose the equivalent problem into two subproblems: one optimizes the transmit multicast and unicast beamformers with a given reflective beamforming, and one optimizes the reflective beamforming with fixed transmit beamformers. For each subproblem, we firstly adopt semidefinite relaxation (SDR) to remove the rank-one constraint to obtain the relaxation solution, and then employ the sequential rank-one constraint relaxation (SROCR) to obtain the beamforming matrices satisfying the rank-one constraint. We further obtain the solution satisfying rank-one constraint via eigenvalue decomposition for each subproblem.
\item Simulation results demonstrate that the proposed iterative algorithm converges in a few iterations and show the impacts of imperfect SIC on the achievable performance. In particular, the proposed scheme outperforms the NOMA-ISAC system under the perfect SIC, while the communication rate achieved by all schemes keeps almost the same with imperfect SIC.
\end{itemize}

\textit{Notations}: Scalars, vectors, and matrices are expressed as lower-case, bold-face, lower-case, and bold-face upper-case letters, respectively; ${\mathbb{C}^{N \times 1}}$ and ${\operatorname{x} ^H}$ represents the space of ${N \!\times\! 1}$ complex-valued vectors and the conjugate transpose of vector $\operatorname{x}$; $\mathcal{C}\mathcal{N}\left( {\mu ,\sigma _c^2} \right)$ denotes the distribution of circularly symmetric complex Guassion (CSCG) random variable with mean $\mu $ and variance ${\sigma _c^2}$; ${\text{rank}}\left( {\mathbf{B}} \right)$, ${\text{Tr}}\left( {\mathbf{B}} \right)$ and ${\text{diag}}\left( {\mathbf{B}} \right)$ represent the rank and trace of matrix ${\mathbf{B}}$ and a vector whose elements are extracted from the main diagonal elements of matrix ${\mathbf{B}}$, respectively; ${\mathbf{B}}  \underline  \succ  0$ indicates that ${\mathbf{B}}$ is a positive semidefinite matrix; ${\mathbb{H}^N}$ and $\mathbb{E}\left(  \cdot  \right)$ represent the set of all $N$-dimensional complex Hermitian matrices and the statistical expectation, respectively.

\section{System Model And Problem Formulation}
\subsection{System Model}
This work considers an IRS-aided NOMA-ISMUC network, which consists of a BS, an IRS, a NU, and a FU, as depicted in Fig. \ref{fig1}. There are two different types of messages sent by the BS, i.e., multicast and unicast signals. Particularly, the multicast signal is for both the NU and FU, while the unicast signal is only for the NU. The BS has $N$ antennas transmitting mixed multicast-unicast signals for communications and sensing with the help of an IRS that has $K$ elements. The channel status information (CSI) is assumed to be perfectly known for us to study the maximum performance gain of the proposed system.

Let ${s_m}\!\left( n \right)\! \sim \mathcal{C}\mathcal{N}\!\left( {0,1} \right)$ and ${s_u}\!\left( n \right) \!\sim \mathcal{C}\mathcal{N}\!\left( {0,1} \right)$ denote the multicast and unicast signasl at time $n$, respectively. Then, the overall transmitted signal of the BS at time $n$th  is given by
\begin{align}\label{1}
{\mathop{\bf x}\nolimits} \left( n \right) = {{\mathop{\bf w}\nolimits} _m}{s_m}\left( n \right) + {{\mathop{\bf w}\nolimits} _u}{s_u}\left( n \right),
\end{align}
where ${\operatorname{\bf w} _m} \!\in\! {\mathbb{C}^{N \times 1}}$ and ${\operatorname{\bf w} _u}\! \in \!{\mathbb{C}^{N \times 1}}$ represent the  beamforming weights for transmitting the multicast and unicast  symbols, respectively.
With the help of an IRS, the received signal of the NU and the FU at  $n$ can be respectively expressed as
\begin{align}\label{2}
\!\!{{\rm{y}}_n}\!\left( n \right) \!\!=\!\! \left( {{\mathop{\bf h}\nolimits} _{I,n}^H{\bf{\Phi G}} \!+\! {\mathop{\bf h}\nolimits} _{B,n}^H} \right){\mathop{\bf x}\nolimits} \left( n \right)\!\! +\! {z_n}\!\!\left( n \right),
\end{align}
\begin{align}\label{3}
\!\!{{\rm{y}}_f}\!\left( n \right)\!\!=\!\! \left( {{\mathop{\bf h}\nolimits} _{I,f}^H{\bf{\Phi G}} \!+\! {\mathop{\bf h}\nolimits} _{B,f}^H} \right){\mathop{\bf x}\nolimits} \left( n \right) \!\! +\!{z_f}\!\!\left( n \right),
\end{align}
where ${\mathbf{G}} \!\! \in\!\! {\mathbb{C}^{K \times N}}$, ${\bf h}_{B,n}^H \!\! \in \!\!{\mathbb{C}^{1 \times N}}$, ${\bf h}_{B,f}^H\!\! \in\!\! {\mathbb{C}^{1 \times N}}$, ${\bf h}_{I,n}^H \!\in\!{\mathbb{C}^{1 \times K}}$ and ${\bf h}_{I,f}^H\!\in \!{\mathbb{C}^{1 \times K}}$ denote the channel matrix/vector of BS-IRS, BS-NU, BS-FU, IRS-NU and IRS-FU links, respectively, ${z_n}\!\left( n \right)\! \!\sim \!\! \mathcal{C}\mathcal{N}\!\left( {0,\sigma _n^2} \right)$ and ${z_f}\!\left( n \right)\! \sim \mathcal{C}\mathcal{N}\!\left( {0,\sigma _f^2} \right)$ are the additive white Gaussian noise (AWGN), and ${\mathbf{\Phi }}{\text{ \!=\! diag}}({e^{j{\theta _1}}}, \cdots ,{e^{j{\theta _K}}})$ denotes the corresponding reflection beamforming matrix with the phase shift of reflecting element ${\theta _k} \!\!\in\!\! \left( {0,2\pi } \right]$.

According to the SIC order considered in the conventional downlink NOMA\cite{9729087}, the multicast signal is firstly detected at the NU and subtracted from the received signal before the unicast signal is decoded. For this reason, the achievable rate for the multicast signal at the NU is written as
\begin{align}\label{4}
{R_{m,N}}\! = \!{\log _2}\left( {1 \!+\! \frac{{{{\left| {\left( {{\mathop{\bf h}}_{I,n}^H{\mathbf{\Phi }}{\mathbf{G}} + {\mathop{\bf h}}_{B,n}^H} \right){\operatorname{\bf w} _m}} \right|}^2}}}{{{{\left| {\left( {{\mathop{\bf h}}_{I,n}^H{\mathbf{\Phi }}{\mathbf{G}} + {\mathop{\bf h}}_{B,n}^H} \right){\operatorname{\bf w} _u}} \right|}^2}\!\! + \!\sigma_n^2}}} \right).
\end{align}
Here we consider an imperfect SIC at the NU, which means that the multicast signal may not be perfectly removed at the NU. Thus, the rate of unicast signals at NU is expressed as
\begin{align}\label{5}
{R_u}\! = \!{\log _2}\left( {1\! + \!\frac{{{{\left| {\left( {{\mathop{\bf h}}_{I,n}^H{\mathbf{\Phi G}}\! + \!{\mathop{\bf h}}_{B,n}^H} \right){\operatorname{\bf w} _u}} \right|}^2}}}{{\zeta {{\left| {\left( {{\mathop{\bf h}}_{I,n}^H{\mathbf{\Phi G}}\! +\! {\mathop{\bf h}}_{B,n}^H} \right){\operatorname{\bf w} _m}} \right|}^2} \!+\! \sigma_n^2}}} \right),
\end{align}
where ${\zeta \!\in\! \left[ {0,1} \right)}$ is a coefficient factor. Please note that ${\zeta \! = \!0}$ and $\zeta \!\!>\!\!0$ correspond to the perfect and imperfect  SIC scenarios, respectively.

As for the FU, the multicast signal is directly detected by treating the unicast signal as interference. As a result, the multicast achievable rate at the FU can be written as
\begin{align}\label{6}
{R_{m,F}}\! = \!{\log _2}\left( {1 \!+ \!\frac{{{{\left| {\left( {{\mathop{\bf h}}_{I,f}^H{\mathbf{\Phi G}} \!+\! {\mathop{\bf h}}_{B,f}^H} \right){\operatorname{\bf w} _m}} \right|}^2}}}{{{{\left| {\left( {{\mathop{\bf h}}_{I,f}^H{\mathbf{\Phi G}}\! +\! {\mathop{\bf h}}_{B,f}^H} \right){\operatorname{\bf w} _u}} \right|}^2} \!+ \!\sigma_f^2}}} \right).
\end{align}
Combining (4) with (6), the rate of the multicast signal can be written as
\begin{align}\label{7}
{R_n}{\text{ = min}}\left\{ {{R_{m,N}},{R_{m,F}}} \right\}.
\end{align}

Similar to\cite{1111444}, we consider that  multicast signals and unicast signals can be totally used for wireless sensing. Furthermore, the sensing performance of the radar system is determined by the signal to interference and noise ratio (SINR) at the FU, which  mainly depends on the target illumination power\cite{1111888}. In this case, the target illumination power arising from the signal transmitted by the BS at the FU can be expressed as
\begin{align}\label{8}
\mathcal{P}\left({\bm{\theta }}\right)=\! \mathbb{E}\left[ {{{\left| {{\mathbf{h}}_f^{\rm H}{\mathbf{x}}} \right|}^2}} \right]\!=\! {\mathbf{h}}_f^{\rm H}{\mathbf{R}}{{\mathbf{h}}_f}\!=\! {\text{Tr}}\left( {{\mathbf{Rh}}_f^{\rm H}{{\mathbf{h}}_f}} \right),
\end{align}
where ${\mathbf{h}}_f^{\rm H} \!=\! {\mathbf{h}}_{I,f}^{\rm H}{\mathbf{\Phi G}} + {\mathbf{h}}_{B,f}^{\rm H}$ is the overall BS-IRS-FU channel vector, ${{\bm{\theta }}} = \left[ {{\theta _1},{\theta _2} \cdots {\theta _k}} \right]$ and ${\mathbf{R}} \!\!=\!\! \mathbb{E}{\left[ {{\mathbf{x}}\left( n \right){\mathbf{x}}{{\left( n \right)}^{\rm H}}} \right]}\!\!\!\! =\!\!\! {{\mathbf{w}}_u}{\mathbf{w}}_u^{\rm H} \!+\!\! {{\mathbf{w}}_m}{\mathbf{w}}_m^{\rm H}$ denotes the corresponding transmit covariance matrix.

\subsection{Problem Formulation}
We aim to maximize the achievable unicast rate at the NU by jointly optimizing the beamformers (${\operatorname{\bf w} _u}$ and ${\operatorname{\bf w} _m}$) for transmitting the multicast and unicast
symbols, and the reflective beamforming (${\mathbf{\Phi }}$) at the IRS. Accordingly, the optimization problem can be formulated as
\begin{align}
\mathop {{\text{max}}}\limits_{{{\operatorname{\bf w} _u}},{\operatorname{\bf w} _m},{\mathbf{\Phi }}}\!\!& {\log _2}\left( {1\! + \!\frac{{{{\left| {\left( {{\mathop{\bf h}}_{I,n}^H{\mathbf{\Phi G}}\! + \!{\mathop{\bf h}}_{B,n}^H} \right){\operatorname{\bf w} _u}} \right|}^2}}}{{\zeta {{\left| {\left( {{\mathop{\bf h}}_{I,n}^H{\mathbf{\Phi G}}\! +\! {\mathop{\bf h}}_{B,n}^H} \right){\operatorname{\bf w} _m}} \right|}^2} \!+\! \sigma_n^2}}} \right)\\
{\text{s}}{\text{.t}}{\text{.}}\;\;\;\;&{R_n} \ge {R} {_m},\tag{9a} \label{9a}\\
&{\text{Tr}}\left( {{\mathop{\bf w}}_u}{\mathop{\bf w}}_u^{\rm H}+{{\mathop{\bf w}}_m}{\mathop{\bf w}}_m^{\rm H} \right)\leq{P_{\max }},\tag{9b}\\
&\mathcal{P}\left( \theta  \right) \ge \Gamma,\tag{9c}
\end{align}
where ${R} {_m}$ denotes the minimum rate requirement of muliticast, ${P_{\max }}$ represents the maximum transmit power budget, and $\Gamma $ is the minimum beam-pattern gain threshold at the FU. (9a) is the minimum rate constraint for the multicast signal, (9b) constrains the maximum   transmit power of the BS, and (9c) constrains the worst-case target illumination power. Problem (9) is non-convex due to the coupled optimization variables i.e., ${{\operatorname{\bf w} _u},{\operatorname{\bf w} _m},{\mathbf{\Phi }}}$ and the fractional form of the objective function.

\section{Proposed Solution}
We note that maximizing ${R_u}$ is equivalent to maximizing the corresponding received SINR. This can be used to simplify the objective function. In what follows, we drop the log function in the objective function of (9). By doing so, the transformed problem  becomes a typical nonlinear programming optimization problem, which can be solved by designing an efficient Dinkelbach-based iterative algorithm \cite{6251827,9103313}. More specifically, we define an equivalent solution for the maximum achievable rate of ${q^ * }$ as:
\begin{align} \label{10}
\begin{split}
{q^ * } &= \frac{{{{\left| {\left( {{\mathop{\bf h}}_{I,n}^H{\mathbf{\Phi G}}\! + \!{\mathop{\bf h}}_{B,n}^H} \right){\operatorname{\bf w} _u}} \right|}^2}}}{{\zeta {{\left| {\left( {{\mathop{\bf h}}_{I,n}^H{\mathbf{\Phi G}}\! +\! {\mathop{\bf h}}_{B,n}^H} \right){\operatorname{\bf w} _m}} \right|}^2} \!+\! \sigma_n^2}}\\
&= \mathop {{\text{max}}}\limits_{{{\operatorname{\bf w} _u}},{\operatorname{\bf w} _m},{\mathbf{\Phi }}} \frac{{{{\left| {\left( {{\mathop{\bf h}}_{I,n}^H{\mathbf{\Phi G}}\! + \!{\mathop{\bf h}}_{B,n}^H} \right){\operatorname{\bf w} _u}} \right|}^2}}}{{\zeta {{\left| {\left( {{\mathop{\bf h}}_{I,n}^H{\mathbf{\Phi G}}\! +\! {\mathop{\bf h}}_{B,n}^H} \right){\operatorname{\bf w} _m}} \right|}^2} \!+\! \sigma_n^2}}.
\end{split}
\end{align}
For the sake of brevity, let $U$ and $M$ denote the numerator and denominator of the fractional expression in (10) respectively, i.e., $U\left( {{\operatorname{\bf w} _u},{\operatorname{\bf w} _m},{\mathbf{\Phi }}} \right)={{{\left| {\left( {{\mathop{\bf h}}_{I,n}^H{\mathbf{\Phi G}}\! + \!{\mathop{\bf h}}_{B,n}^H} \right){\operatorname{\bf w} _u}} \right|}^2}}$, $M\left( {\operatorname{\bf w} _u,\operatorname{\bf w} _m,{{\mathbf{\Phi }}}} \right)={{\zeta {{\left| {\left( {{\mathop{\bf h}}_{I,n}^H{\mathbf{\Phi G}}\! +\! {\mathop{\bf h}}_{B,n}^H} \right){\operatorname{\bf w} _m}} \right|}^2} \!+\! \sigma_n^2}}$. Applying the generalized fractional programming theory, we have the following Theorem \cite{6251827,9103313}.

\textbf{Theorem.} The equivalent maximum achievable rate $q^ *$ can be achieved if and only if
\begin{align}\label{11}
\begin{split}
&\mathop {{\text{max}}}\limits_{{{\operatorname{\bf w} _u}},{\operatorname{\bf w} _m},{\mathbf{\Phi }}} {\text{      }} U\left( {{\operatorname{\bf w} _u},{\operatorname{\bf w} _m},{\mathbf{\Phi }}} \right) - {q^*}M\left( {\operatorname{\bf w} _u,\operatorname{\bf w} _m,{{\mathbf{\Phi }}}} \right)\\ &= U\left( {\operatorname{\bf w} _u^*,\operatorname{\bf w} _m^*,{{\mathbf{\Phi }}^*}} \right) - {q^*}M\left( {\operatorname{\bf w} _u^*,\operatorname{\bf w} _m^*,{{\mathbf{\Phi }}^*}} \right) \\&=  0,
\end{split}
\end{align}
where * is the optimal solution of the optimization variable.

Using this Theorem, we propose Algorithm 1 to solve problem (9), as shown at the top of the next page, where the main challenge is to solve the following problem (12),
\begin{align}
\mathop {{\text{max}}}\limits_{{{\operatorname{\bf w} _u}},{\operatorname{\bf w} _m},{\mathbf{\Phi }}}\;\;& U\left( {{\operatorname{\bf w} _u},{\operatorname{\bf w}_m},{\mathbf{\Phi }}} \right) - {q}M\left( {{\operatorname{\bf w}_u},{\operatorname{\bf w}_m},{\mathbf{\Phi }}}\right)\\
{\text{s}}{\text{.t}}{\text{.}}\;\;\;\;&{{\text{(9a)--(9c)}}}.\tag{12a} \label{12a}
\end{align}

Although the equivalent objective function in (12) is more tractable than the original optimization problem, the coupling between the transmit beamformers and the reflective beamforming still exists. We  employ the alternating optimization algorithm to decompose the original problem into two subproblems, where the transmit beamformers (${\operatorname{\bf w} _u}$ and ${\operatorname{\bf w} _m}$) at the BS and the reflective beamforming (${\mathbf{\Phi }}$) at the IRS could be alternatively optimized.
\begin{algorithm}
\caption{The iterative allgorithm to solve problem (9)}
\label{alg:algorithm1}
\begin{algorithmic}
\STATE Initialize: the maximum number of iteration $L$ and the\\ maximum convergence threshold ${ \mathrel\backepsilon  _1}$,\\
Set maximum achievable rate $q\! = \!0$ and iteration index $i\!=\!0$
\REPEAT
\STATE Solve problem (12) for a given $q$ and obtain\! $\left\{\!{{{\mathbf{W}}_u},\!{{\mathbf{W}}_m},\!{\!\mathbf{\Phi }}}\! \right\}$
\IF{$U\left( {{\operatorname{\bf w} _u},{\operatorname{\bf w} _m},{\mathbf{\Phi }}} \right) - {q}M\left( {\operatorname{\bf w} _u,\operatorname{\bf w} _m,{{\mathbf{\Phi }}}} \right) < { \mathrel\backepsilon  _1}$ }
\STATE Convergence = \textbf{true}
\RETURN $\left\{{\operatorname{\bf w} _u^*,\operatorname{\bf w} _m^*,{{\mathbf{\Phi }}^*}}\right\}=\left\{ {{{\mathbf{w}}_u},{{\mathbf{w}}_m},{\mathbf{\Phi }}} \right\}$ and \\
${{{q}}^*} = \frac{{{{ U}}\left( {{\operatorname{\bf w} _u},{\operatorname{\bf w} _m},{\mathbf{\Phi }}} \right)}}{{{{M }}\left( {{\operatorname{\bf w} _u},{\operatorname{\bf w} _m},{\mathbf{\Phi }}} \right)}}$
\ELSE
\STATE  Set ${{{q}}} = \frac{{{{ U}}\left( {{\operatorname{\bf w} _u},{\operatorname{\bf w} _m},{\mathbf{\Phi }}} \right)}}{{{{M }}\left( {{\operatorname{\bf w} _u},{\operatorname{\bf w} _m},{\mathbf{\Phi }}} \right)}}$ and $i=i+1$\vspace{0.1cm}\\Convergence = \textbf{false}
\ENDIF
\vspace{-0.2cm}
\STATE
\UNTIL Convergence = \textbf{true} or $i=L$
\end{algorithmic}
\end{algorithm}

\subsection{Updating ${\operatorname{\bf w} _u}$ and ${\operatorname{\bf w} _m}$ given ${\mathbf{\Phi }}$}
First, we aim to optimize the transmit beamformers ${\operatorname{\bf w} _u}$ \!and ${\operatorname{\bf w} _m}$\! in problem (12) with a given \!${\mathbf{\Phi }}$, i.e.,
\begin{align}\label{12}
\mathop {{\text{max}}}\limits_{{\operatorname{\bf w} _u},{\operatorname{\bf w} _m}} &{\text{    }}{\left| {{\mathop{\bf h}}_n^H{\operatorname{\bf w} _u}} \right|^2} - q\left( {\zeta {{\left| {{\mathop{\bf h}}_n^H{\operatorname{\bf w} _m}} \right|}^2} + \sigma_n^2} \right)\\
{\text{s.}}{\text{t.}}\;\;&{\text{(9a)--(9c)}},\tag{13a}\label{13a}
\end{align}
where ${\mathop{\bf h}}_n^{\rm H} \!\!= \!\!\left( {{\mathop{\bf h}}_{I,n}^{\rm H}{\mathbf{\Phi G}} \!+\! {\mathop{\bf h}}_{B,n}^{\rm H}} \right)$ represents the  combined channel from BS-IRS-NU. We define ${{\mathbf{W}}_m} \!\!=\!\! {\operatorname{\bf w} _u}{\operatorname{\bf w} _u}^{\rm H}$ and ${{\mathbf{W}}_m} \!\!=\!\! {\operatorname{\bf w} _m}{\operatorname{\bf w} _m}^{\rm H}$, ${{\mathbf{W}}_m}\underline  \succ  {\text{ 0}}$, ${{\mathbf{W}}_m}\underline  \succ  {\text{ 0}}$, ${\text{rank(}}{{\mathbf{W}}_m}{\text{) = 1}}$, and ${\text{rank(}}{{\mathbf{W}}_u}{\text{) = 1}}$.  The problem (13) can be rewritten as
\begin{align}\label{14}
\mathop{\max}\limits_{{\mathbf{W}}{_u},{\mathbf{W}}{_m}}&{\text{     Tr}}({{\mathbf{H}}_n}{{\mathbf{W}}_u}) - q(\zeta {\text{     Tr}}({{\mathbf{H}}_n}{{\mathbf{W}}_m}) + \sigma _n^2) \\
{\text{s.}}{\text{t.}}\;\;&{\text{Tr}}\left( {\mathbf{R}} \right)\leq{P_{\max }},\tag{14a}\label{14a}\\
&{\text{Tr}}({{\mathbf{H}}_n}{{\mathbf{W}}_m}) -{ \overline \gamma}{_m} {\text{Tr}}({{\mathbf{H}}_n}{{\mathbf{W}}_u}) - {\overline \gamma}{_m} \sigma _n^2 \ge 0,\tag{14b}\label{14b}\\
&{\text{Tr}}({{\mathbf{H}}_f}{{\mathbf{W}}_m}) -{\overline \gamma}{_m} {\text{Tr}}({{\mathbf {H}}_f}{{\mathbf{W}}_u}) - {\overline \gamma}{_m} \sigma _f^2 \ge 0, \tag{14c}\label{14c}\\
&{{\mathbf{W}}_u},{{\mathbf{W}}_m}\underline  \succ  {\text{ 0,}}{{\mathbf{W}}_u},{{\mathbf{W}}_m} \in {\mathbb{H}^N},\tag{14d}\label{14d}\\
&{\text{Tr}}\left( {\mathbf{Rh}}_f^{\rm H}{{\mathbf{h}}_f} \right) \ge \Gamma, \tag{14e}\label{14e}\\
&{\text{rank(}}{{\mathbf{W}}_u}{) = 1,\text{rank(}}{{\mathbf{W}}_m}{) = 1},\tag{14f}\label{14f}
\end{align}
where ${\mathop{\bf h}}_f^{\rm H}\! =\!\! \left( {{\mathop{\bf h}}_{I,f}^{\rm H}{\mathbf{\Phi G}}\! +\! {\mathop{\bf h}}_{B,f}^{\rm H}} \right)$ denotes the overall channel vector from BS-IRS-FU, ${\overline \gamma}{_m}\!\! = \!\!{2^{ {{R}}_m }}\!\! - \!\!1$, and (14a) and (14b) are derived from \eqref{9a}. In this problem, constraint (14f) is non-convex, which motivates us to use the semidefinite relaxation (SDR) relaxing the rank-one constraint to obtain the relaxation solution, and then to adopt the Sequential Rank-One Constraint Relaxation (SROCR)\cite{8081370} to obtain beamforming matrices satisfying rank-one constraint. Thus, problem (14) would be reformulated as
\begin{align} \label{15}
\mathop {\max }\limits_{{\mathbf{W}}{_u},{\mathbf{W}}{_m}}& {\text{     }}{\text{Tr}}({{\mathbf{H}}_n}{{\mathbf{W}}_u}) - q(\zeta {\text{Tr}}({{\mathbf{H}}_n}{{\mathbf{W}}_m}) + \sigma_n^2) \\
{\text{s}}{\text{.t}}{\text{.       }}&\;\;{{\mathbf{u}}^H}_{{{\max }_u}}{{\mathbf{W}}_u}{{\mathbf{u}}_{{{\max }_u}}} \ge m_u^{\left( i \right)}{\text{Tr}}\left( {{{\mathbf{W}}_u}} \right),\tag{15a}\label{15a}\\
&\;\;{{\mathbf{u}}^H}_{{{\max }_m}}{{\mathbf{W}}_m}{{\mathbf{u}}_{{{\max }_m}}} \ge m_m^{\left( i \right)}{\text{Tr}}\left( {{{\mathbf{W}}_m}} \right),\tag{15b}\label{15b}\\
&\;\;{\text{(14a)--(14e)}},\tag{15c}\label{15c}
\end{align}
where ${{\mathbf{u}}_{{{\max }_u}}}$ and ${{\mathbf{u}}_{{{\max }_m}}}$ denote the respective largest eigenvalues of ${{\mathbf{W}}_u}$ and ${{\mathbf{W}}_m}$, and $m_u^{\left( i \right)}$ and $m_m^{\left( i \right)}$ are the relaxation parameters controlling the largest eigenvalue to trace ratio of ${{\mathbf{W}}_u}$ and ${{\mathbf{W}}_m}$, respectively.
We transfer problem (13) into problems (14) and (15) to obtain the optimization solutions. Specifically, to tackle the rank-one constraint, we firstly drop the constraint (14f) in problem (14) to get an SDR problem, which is convex and can be solved by using existing standard convex problem solvers such as CVX \cite{Cvex}. We than settle the convex problem (15) with SROCR algorithm via CVX solvers to obtain the beamforming matrices satisfying rank-one constraint.
Based on the above analysis, the proposed algorithm is summarized in the Algorithm 2.
\begin{algorithm}
\caption{The Proposed Algorithm for problem (13)}
\label{alg:algorithm2}
\begin{algorithmic}
\STATE Initialize: $i = 0$, convergence threshold ${ \mathrel\backepsilon  _2}$,\\Solve the problem(14) without (14f) and obtain \!${\mathbf{W}}_u^{\left( 0 \right)}$\! and \!${\mathbf{W}}_m^{\left( 0 \right)}$\!, while $m_u^{\left( 0 \right)} \!\!= 0$ \\and $m_m^{\left( 0 \right)} \!\!= 0$.\\Define the initial step size:
\begin{align}\notag
\delta_{e}^{(0)} \in\left(0,1-\frac{\lambda_{\max }\left(\mathbf{W}_{e}^{(0)}\right)}{\text{Tr}\left(\mathbf{W}_{e}^{(0)}\right)}\right],{\text{e}} \in \left( {u,m} \right).
\end{align}
\REPEAT
\STATE Given $ {{{\mathbf{W}}_u}^{\left( i \right)},{m_u}^{\left( i \right)}} $ and $ {{{\mathbf{W}}_m}^{\left( i \right)},{m_m}^{\left( i \right)}} $,\\solve the problem (15)
\IF{Problem (15) is solvable }
\STATE Obtain the optimal solution\!\! ${\mathbf{W}}_u^{\left( {i + 1} \right)}$\!\! and \!\!${\mathbf{W}}_m^{\left( {i + 1} \right)}$ \\for Problem(15); ${\delta _e}^{\left( {i + 1} \right)} \leftarrow {\delta ^{\left( 0 \right)}},{\text{e}} \in \left( {u,m} \right).$
\ELSE
\STATE  \begin{align}\notag
{\delta _e}^{\left( {i + 1} \right)} = {\delta _e}^{\left( i \right)}/3,{\text{e}} \in \left( {u,m} \right).\vspace{-0.5cm}
\end{align}
\ENDIF
\STATE
\begin{align}\notag
m_{e}^{(i+1)}\!\! \leftarrow \!\min\!\! \left(1, \frac{\lambda_{\max }\left(\mathbf{W}_{e}^{(i+1)}\right)}{\text{Tr}\left(\mathbf{W}_{e}^{(i+1)}\right)}\!+\!\delta_{e}^{(i+1)}\!\!\right)\!,{\text{e}} \in \left( {u,m} \right);
\end{align}
\UNTIL{${m_e}^{\left( {i - 1} \right)} \ge { \mathrel\backepsilon  _2}$}
\end{algorithmic}
\end{algorithm}

\subsection{Updating ${\mathbf{\Phi }}$ given ${{{\mathop{\bf w}}_u}}$ and ${{{\mathop{\bf w}}_m}}$}
Next, we optimize the reflective beamforming ${\mathbf{\Phi }}$ with the fixed transmit beamformers \!(${\operatorname{\bf w} _u}$ and ${\operatorname{\bf w} _m}$), i.e.,
\begin{align}\label{15}
\max _{{\mathbf{\Phi }}}&\left|\left({\mathbf{\Phi }} \mathbf{H}_{I, n}\!\!+\!\!\mathrm{\bf h}_{B, n}^{H}\right) \!\mathrm{\bf w}_{u}\right|^{2}\!\!\!-\!q\left(\zeta \!\left|\left({\mathbf{\Phi }} \mathbf{H}_{I, n}\!\!+\!\!\mathrm{\bf h}_{B, n}^{H}\right) \mathrm{\bf w}_{m}\right|^{2}\!\!\!+\!\sigma_{n}^{2}\right)\\
{\text{s.}}{\text{t.}}&\;\; {\text{(9a),(9c)}},\tag{16a}\label{16a}
\end{align}
where ${{\mathbf{H}}_{I,n}}\! = \!{\text{diag}}({\bf h}_{I,n}^H){\mathbf{G}}$, and ${{\mathbf{H}}_{I,f}} \! = \! {\text{diag}}({\bf h}_{I,f}^H){\mathbf{G}}$. On this basis, we have ${\mathop{\bf v}}{\text{ = [}}{e^{j{\theta _1}}}, \cdots ,{e^{j{\theta _N}}}{{\text{]}}^{\rm H}}$ denoting the reflective phase shift vector at the IRS, and define the following variables
\begin{align}\label{16}
\overline {\mathbf{v}} {\text{ = }}{\left[ {{\mathbf{v}},1} \right]^{\rm H}},\;\;\;\;\;\;{\mathbf{Z}} = \operatorname{diag} \left( {{\mathbf{h}}_{I,f}^{\rm H}} \right),
\end{align}
\begin{align}\label{17}
{{\mathbf{F}}}={
\left[ \begin{array}{cc}
{{\mathbf{Z}}^{\rm H}}{\mathbf{GR}}{{\mathbf{G}}^{\rm H}}{\mathbf{Z}}
&{{\mathbf{Z}}^{\rm H}}{\mathbf{GR}}{{\mathbf{h}}_{B,f}} \\
{\mathbf{h}}_{B,f}^{\rm H}{\mathbf{R}}{{\mathbf{G}}^{\rm H}}{\mathbf{Z}}
& {\mathbf{h}}_{B,f}^{\rm H}{\mathbf{R}}{{\mathbf{h}}_{B,f}}
\end{array}
\right ]},
\end{align}
\begin{align}\label{18}
{{\mathbf{R}}_{ij}}\!=\!\!{
\left[ \begin{array}{cc}
\!\!{{\mathbf{H}}_{I,i}}{{\mathbf{W}}_j}{\mathbf{H}}_{I,i}^{\rm H}  & \!\! {{\mathbf{H}}_{I,i}}{{\mathbf{W}}_j}{{\text{h}}_{B,i}}\!\! \\
\!\!{\text{h}}_{B,i}^H{{\mathbf{W}}_j}{\mathbf{H}}_{I,i}^{\rm H}    &   {\mathbf{0}}\!\!
\end{array}
\right ]}, \left( \begin{gathered}
  i \in \left( {n,f} \right) \hfill \\
  j \in \left( {u,m} \right) \hfill \\
\end{gathered}  \right).
\end{align}
Then, using (17) and (18), the target illumination power at the FU can be rewritten as
\begin{align}\label{18}
\mathcal{P}\left( \theta  \right) \!=\! \mathbb{E}\left[ {{{\left| {{\mathbf{h}}_f^{\rm H}{{\mathbf{x}}}} \right|}^2}} \right]\!\!=\!\!\left|\left[\mathbf{G}^{\mathrm{H}} \mathbf{Z},\!\!\!\! \quad \mathbf{h}_{B, f}\right] \mathbf{x}\right|^{2} \!\!= \!{\overline {\mathbf{v}} ^{\rm H}}{\mathbf{F}}\overline {\mathbf{v}}.
\end{align}
Thus, problem (16) is equivalent to
\begin{align}\label{21}
\mathop {\max }\limits_{\overline {\mathbf{v}}}&\;\;{\text{Tr}}({\overline {\mathbf{v}} ^{\rm H}}{{\mathbf{R}}_{nu}}\overline {\mathbf{v}} )\! + \!{{\text{A}}_{nu}} \!
-\! q(\zeta {\text{Tr}}({\overline {\mathbf{v}} ^{\rm H}}{{\mathbf{R}}_{nm}}\overline {\mathbf{v}} )\! +\! {{\text{A}}_{nm}})\! +\! \sigma _n^2) \\
{\text{s.}}{\text{t.}}&\;\;{\text{Tr}}({\overline {\mathbf{v}} ^{\rm H}}{{\mathbf{R}}_{nm}}\overline {\mathbf{v}} ) - {\overline \gamma  _m}{\text{Tr}}({\overline {\mathbf{v}} ^{\rm H}}{{\mathbf{R}}_{nu}}\overline {\mathbf{v}}) + {\text{C}} \ge 0,\tag{21a} \label{21a}\\
&\;\;{\text{Tr}}({\overline {\mathbf{v}} ^{\rm H}}{{\mathbf{R}}_{fm}}\overline {\mathbf{v}} ) - {\overline \gamma  _m}{\text{Tr}}({\overline {\mathbf{v}} ^{\rm H}}{{\mathbf{R}}_{fu}}\overline {\mathbf{v}}) + {\text{D}}\ge 0, \tag{21b}\label{21b}\\
&\;\; {\overline {\mathbf{v}} ^{\rm H}}{\mathbf{F}}\overline {\mathbf{v}} \ge \Gamma,\tag{21c}\label{21c}
\end{align}
where (21a) and (21b) are derived from (9a), and (21c) is the constraint of the worst target illumination power. Let
${{\text{A}}_{nu}}\!\!= \!\!{\mathop{\bf h}}_{B,n}^H{{\mathbf{W}}_u}{{\mathop{\bf h}}_{B,n}}$, ${{\text{A}}_{nm}}\!\! = \!\! {\mathop{\bf h}}_{B,n}^H{{\mathbf{W}}_m}{{\mathop{\bf h}}_{B,n}}$, ${\text{C}} = {\mathop{\bf h}}_{B,n}^H{{\mathbf{W}}_m}{{\mathop{\bf h}}_{B,n}} - {\overline \gamma  _m}({\mathop{\bf h}}_{B,n}^H{{\mathbf{W}}_u}{{\mathop{\bf h}}_{B,n}}) \!-\! {\overline \gamma  _m}\sigma _n^2$  and ${\text{D} =\text{ h}}_{B,f}^H{{\mathbf{W}}_m}{{\mathop{\bf h}}_{B,f}} - {\overline \gamma _m}({\mathop{\bf h}}_{B,f}^H{{\mathbf{W}}_u}{{\mathop{\bf h}}_{B,f}}) - {\overline \gamma  _m}\sigma _f^2$ for the sake of brevity. It is obvious that the optimization of ${\mathbf{\Phi }}$ is replaced by the optimization of $\overline {\mathop{\bf v}}$. Define ${\mathbf{V}} = \overline {\mathop{\bf v}} \!{\text{ }}{\overline {\mathop{\bf v}} ^{\rm H}}$, which satisfies ${\mathbf{V}}\underline  \succ  0$ and ${\text{rank}}({\mathbf{V}}) \!=\! 1$. Thus, problem (21) is reformulated as
\begin{align}\label{22}
\mathop {\max }\limits_{\mathbf{V}}&{\text{ }}{\text{Tr}}({{\mathbf{R}}_{nu}}{\mathbf{V}})\! + \!{{\text{A}}_{nu}} \!
-\! q(\zeta ({\text{Tr}}({{\mathbf{R}}_{nm}}{\mathbf{V}})\! +\! {{\text{A}}_{nm}})\! +\! \sigma _n^2) \\
{\text{s.}}{\text{t.}}&\;\;{\text{Tr}}({{\mathbf{R}}_{nm}}{\mathbf{V}}) - {\overline \gamma  _m}{\text{Tr}}({{\mathbf{R}}_{nu}}{\mathbf{V}}) + {\text{C}} \ge 0,\tag{22a} \label{22a}\\
&\;\;{\text{Tr}}({{\mathbf{R}}_{fm}}{\mathbf{V}}) - {\overline \gamma  _m}{\text{Tr}}({{\mathbf{R}}_{fu}}{\mathbf{V}}) + {\text{D}}\ge 0, \tag{22b}\label{22b}\\
&\;\;{\text{Tr}}\left( {{{\mathbf{F}}}{\mathbf{V}}} \right) \ge \Gamma ,\tag{22c}\label{22c}\\
&\;\;{{\mathbf{V}}_{n,n}}{\text{ = 1, }}\forall {\text{n}} \in \left\{ {1,2,...,N + 1} \right\}, {\mathbf{V}}\underline  \succ  0,\tag{22d}\label{22d}\\
&\;\;{\text{rank}}({\mathbf{V}}) = 1.\tag{22e}\label{22e}
\end{align}
Accordingly, problem (22) using SROCR algorithm can
be reformulated as
\begin{align}
\mathop {\max }\limits_{\mathbf{V}}&{\text{ }}{\text{Tr}}({{\mathbf{R}}_{nu}}{\mathbf{V}})\! + \!{{\text{A}}_{nu}} \!
-\! q(\zeta ({\text{Tr}}({{\mathbf{R}}_{nm}}{\mathbf{V}})\! +\! {{\text{A}}_{nm}})\! +\! \sigma _n^2) \\
{\text{s.}}{\text{t.}}&\;\;{{\mathbf{u}}^H}_{{{\max }_v}}{{\mathbf{V}}}{{\mathbf{u}}_{{{\max }_v}}} \ge m_v^{\left( i \right)}{\text{Tr}}\left( {{{\mathbf{V}}}} \right),\tag{23a}\label{23a}\\
&\;\;{\text{(22a)--(22d)}}.\tag{23b}\label{23b}
\end{align}
Note that we employ the eigenvalue decomposition to obtain the solution $\overline v $ of problem (23), and then we can get the pre-optimize variable $v$ by taking the ${N \times 1}$ elements of $\overline v $. Until now, problem (9) can be solved via repeating the following two steps till convergence: (a) fix reflective beamforming and update the transmit beamformers by solving (13), and (b) update reflective beamforming by solving (16), where the alternating optimization based algorithm for the optimization problem (9) is completed. The iterations required for the convergence of ${q^ * }$ and the alternating algorithm are denoted by ${\Delta _1}$ and ${\Delta _2}$, respectively. For SROCR algorithm solving problem (15) and (23), the interior point method \cite{Cvex} is adopted and its computation complexity are ${\rm \textit{O}}(\sqrt {{I_1}} \log ({I_1}))$ and ${\rm \textit{O}}(\sqrt {{I_2}} \log ({I_2}))$, where $I_1$ and $I_2$ denote the number of the inequality constraints of problem (15) and (23), respectively. Thus, the total computation complexity of the proposed algorithm is ${\rm \textit{O}}\left( {{\Delta _1}{\Delta _2}\left( {\sqrt {{I_1}} \log ({I_1}) + \sqrt {{I_2}} \log ({I_2})} \right)} \right)$.

\section{Simulation Results}
In this section, numerical results are presented to validate the performance of the proposed algorithm. In particular, Rician fading is considers for the BS-IRS, IRS-users and BS-RU links. The BS adopts a uniform linear array (ULA) with half-wavelength spacing between adjacent antennas. The BS-NU is assumed to adopting Rayleigh channel model and the path loss is ${L_{n}}\! =\! {L_0}\! +\! 30{\log _{10}}{d_n}$, while the BS-FU has the virtual line-of-sight (LoS) related with the path loss of ${L_{f}} \!=\!{L_0}\! +\! 20{\log _{10}}{d_f}$. In general, $L_0$ is the path loss at the reference $d\!=\!1$, ${d_f}$ and ${d_n}$ are the distance between the BS and users. We set the number of antennas at the BS and the IRS elements as $K\!=\!14$, $N\!=\!20$, respectively. In particular, the rank-one constrains to 0.99, and the target illumination power, i.e., $\Gamma {\text{ = 1}}{{\text{0}}^{ - 2}}$. The other basis simulation parameters we set are as follows: ${L_0} = 40{\text{dB}}$, ${d_f} = 1000{\text{m}}$, ${d_n} = 100{\text{m}}$, ${{\text{P}}_{\max }} = 10{\text{dBm}}$, and ${\sigma ^2} =  - 100{\text{dBm}}$.
\begin{figure}
  \centering
  \includegraphics[width=0.40\textwidth]{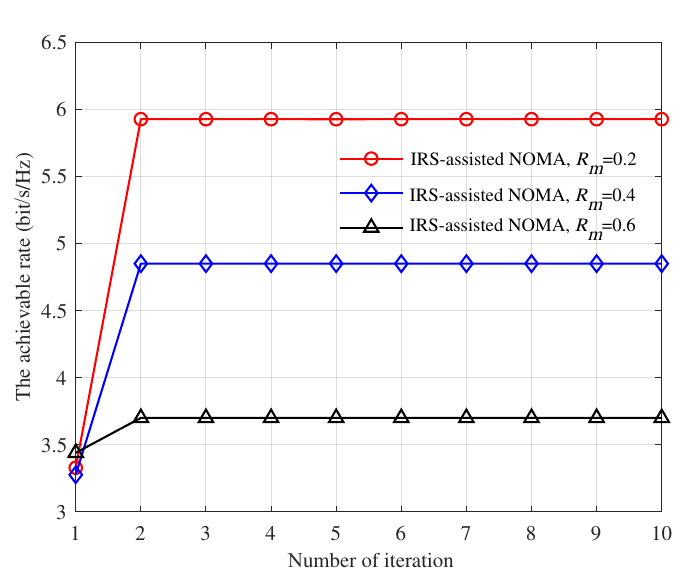}\\
  \caption{The iteration number of the proposed algorithm.}\label{fig2}
\end{figure}

\begin{figure*}[t]
\begin{minipage}[t]{0.33\linewidth}
\centering
\includegraphics[height=5.4cm,width=6cm]{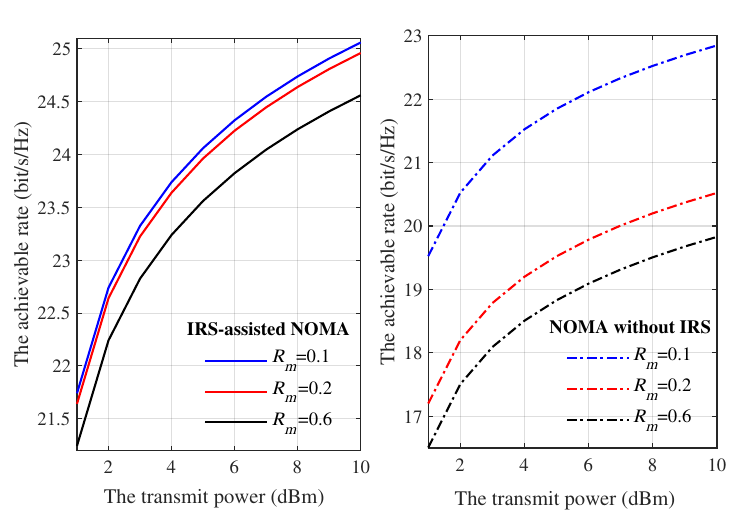}
\captionsetup{font={scriptsize}}\caption{The achievable rate versus ${P_{\text{max}}}$ for different ${R}_{m}$ under perfect SIC.}\label{fig3}
\end{minipage}
\begin{minipage}[t]{0.33\linewidth}
\centering
\includegraphics[height=5.4cm,width=5.6cm]{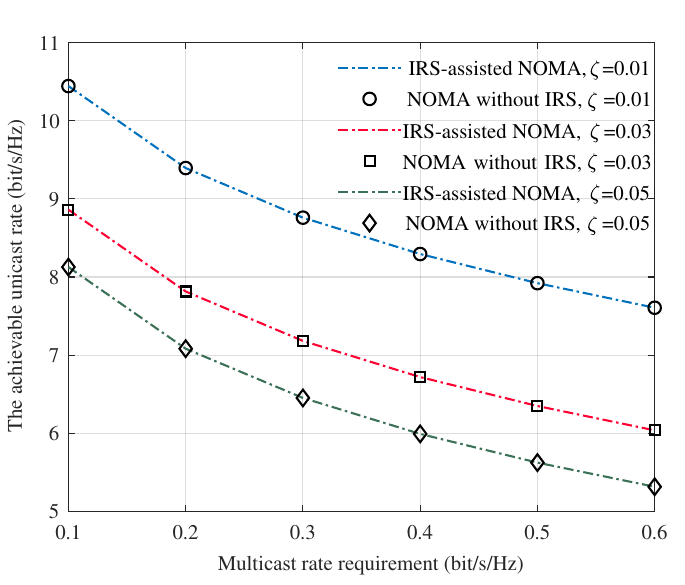}
\captionsetup{font={scriptsize}}\caption{The achievable rate versus ${R}_{m}$ fordifferent\\ $\zeta$under imperfect SIC.}\label{fig4}
\end{minipage}
\begin{minipage}[t]{0.33\linewidth}
\centering
\includegraphics[height=5.4cm,width=5.6cm]{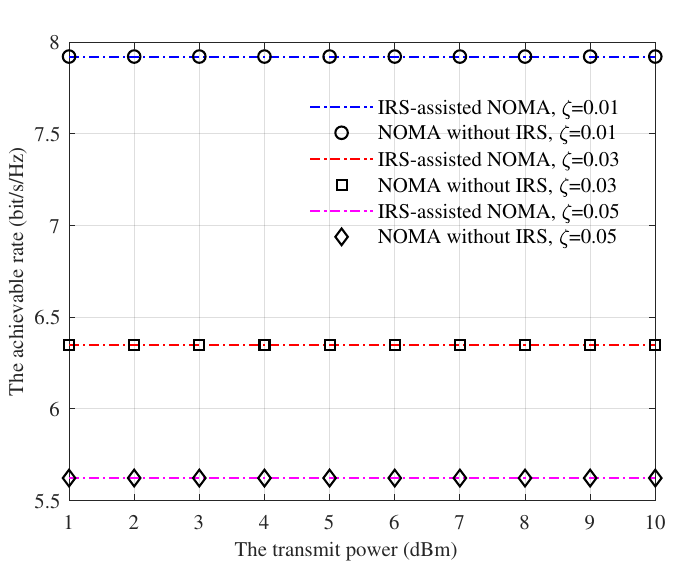}
\captionsetup{font={scriptsize}}\caption{The achievable rate versus ${P_{\text{max}}}$ under imperfect SIC.}\label{fig5}
\end{minipage}%
\end{figure*}



Fig. \ref{fig2} illustrates the convergence behavior of our proposed algorithm versus different multicast rate requirements. It can be observed that the  proposed algorithm always converges within a few iterations, thus validating the effectiveness of the proposed algorithm.
Fig. \ref{fig3} examines the impact of ${P_{\text{max}}}$ on the achievable unicast rate by using different schemes with perfect SIC.
We can see that the unicast achievable rate of all the schemes increases with the increase of the transmit power. This is because that a stronger transmit signal strength for all the users can be obtained with a higher transmit power. We can also observe that a higher ${R} _{m}$ leads to lower unicast rates with a fixed transmit power. By comparing the simulation results in Fig \ref{fig3}, we can conclude that the unicast rate obtained by the proposed scheme achieves a better performance than that of the benchmark scheme.

Fig. \ref{fig4} shows the effect of ${R}_{m}$ on the achievable rate.  From Fig. \ref{fig4}, the unicast rate of all the schemes increases upon decreasing $\zeta$ because a bigger imperfect coefficient embodies more residual interference in a imperfect SIC scenario, which indicates that the interference of the NU is inevitably magnified. We can also observe that the unicast rate achieved by all the schemes decreases as ${R} _{m}$ increases. This is because a higher multicast rate requires more transmit power to be allocated for all the users, thus degrading the unicast rate achieved. Different from the case with perfect SIC, the unicast rate achieved by all the schemes remains almost unchanged in the imperfect SIC scenario. The reason  is that the multicast signal may not be perfectly removed at the NU by employing of imperfect SIC.

Fig. \ref{fig5} presents the achievable rate versus ${P_{\text{max}}}$ with imperfect SIC. As is emerged from Fig. \ref{fig5},  the unicast rate of all the schemes decreases upon increasing $\zeta$. We can also observe that the unicast rate achieved by all the schemes remains unchanged with increasing ${P_{\text{max}}}$. Similar to the analysis for Fig. \ref{fig4}, the transmit signal strength is enhanced as the transmit power increases, while the interference power of the NU increases as well. In  this circumstance, IRS embodies little assistance. The above results also confirm that the performance of IRS-assisted NOMA does not ideally outperform performance of NOMA without IRS.

\section{Conclusions}
In this work, we considered IRS in a NOMA-ISMUC network with perfect/imperfect SIC. We formulated an optimization problem to maximize the unicast rate by jointly optimizing the transmit beamforming at the BS and the reflective beamforming at the IRS, while ensuring a given target illumination power.
Using the techniques of Dinkelbach method, SROCR and alternating optimization, we proposed an efficient
iterative algorithm to solve the formulated problem.
The numerical results verify the fast convergence of the proposed algorithm. The results also show that the proposed scheme outperforms the benchmark scheme with perfect SIC, but the unicast rates achieved by all the schemes are almost the same with imperfect SIC.

\ifCLASSOPTIONcaptionsoff
  \newpage
\fi
\bibliographystyle{IEEEtran}
\bibliography{refa}

\end{document}